\documentclass[doublecol]{epl2} \usepackage{graphicx}

\def\(({\left(} \def\)){\right)}

\newcommand{\Tr}{{\rm Tr}} \newcommand{\be}{\begin{equation}}
  \newcommand{\ee}{\end{equation}} \newcommand{\bea}{\begin{eqnarray}}
  \newcommand{\eea}{\end{eqnarray}} 
\newcommand{\eps}{\varepsilon} \newcommand{\s}{\sigma}
 \newcommand{\e}{\tx {e}}

  \def\I{{\rm
    1\hspace{-0.90ex}1}}

\title{Energy gaps in quantum first-order mean-field-like transitions: \\
The problems that quantum annealing cannot solve}
\author{ T. J\"org$^1$, F. Krzakala$^{2,3}$, J. Kurchan$^4$,
  A. C. Maggs$^2$ and J. Pujos$^2$}
\shortauthor{ T. J\"org, F. Krzakala, J. Kurchan, A. C. Maggs and
  J. Pujos}

\institute{ $^1$ CNRS et ENS UMR 8549, 24 Rue Lhomond, 75231
  Paris Cedex 05, France, LPTENS \\
  $^2$ CNRS; ESPCI ParisTech, 10 rue Vauquelin, UMR 7083 Gulliver,
  Paris, France 75005, PCT \\
  $^3$ Theoretical Division and Center for Nonlinear Studies, Los
  Alamos National Laboratory, NM 87545 USA \\
  $^4$ CNRS; ESPCI ParisTech, 10 rue Vauquelin, UMR 7636, Paris,
  France 75005, PMMH }

\abstract{We study first-order quantum phase transitions in models
  where the mean-field traitment is exact, and the exponentially fast
  closure of the energy gap with the system size at the transition.
  We consider exactly solvable ferromagnetic models, and show that
  they reduce to the Grover problem in a particular limit. We compute
  the coefficient in the exponential closure of the gap using an
  instantonic approach, and discuss the (dire) consequences for
  quantum annealing.}


\begin{document}

\maketitle

Many important practical problems involve the minimization of a
function of discrete variables. Solving such combinatorial problems by
temperature annealing is a classical strategy in computer science
\cite{Scott}: the idea is to use thermal fluctuations to avoid trapping
the system in local minima, and thereby efficiently visit the whole
configuration space. It has been proposed to extend this approach to
quantum fluctuations\cite{QA}; it is thus of interest to ask
whether annealing by tuning down the amplitude of a quantum mechanical
kinetic operator such as a transverse magnetic field $\Gamma$ can
outperform the classical approach. In particular, can problems that
normally take exponential time be solved in only polynomial time?

Some considerable effort has been devoted to this question in the
context of difficult combinatorial problems (see for instance
\cite{Peter}) which have a counterpart in statistical physics where
they corresponds to mean-field spin-glass models \cite{MPV,BOOK}.
However, most of the studies were purely numerical and thus restricted
to very small sizes due to the difficulty of simulating quantum
mechanics without a quantum computer. In a recent Letter \cite{US-PRL}
(see also \cite{US-Long}), we argued that with the usual
implementation of the quantum annealing it is likely that the most
difficult systems undergo a quantum transition of {\it the first
  order} as the transverse field is tuned; this is a generic feature
for many quantum spin glasses\cite{FirstOrder}.  More recently, a
first order transition has indeed been indentified in the phase
diagram of one of the most studied random optimization problems,
called XORSAT \cite{XORSAT}. As we shall see, this implies the failure
of quantum annealing for the hardest optimization problems.

The reason why quantum annealing is {\it not} an efficient strategy
for finding the ground state across a first-order transition can be
understood from a simple argument. Quantum annealing could {\it in
  principle} be more efficient than thermal annealing for certain
classes of problems: From the WKB approximation it is well known that
a quantum particle tunnels rapidly through very high (in energy) but
thin (in distance) energy barriers. Thermal annealing is much better
at low, but deep barrier crossing.  However, in a first-order
transition the two states whose free energies cross are generally {\it
  far} from each other in the phase space; quantum tunneling must be
inefficient. 

To make this argument more precise, and to compute how slow an
annealing should be such that the tunelling do happens, one can
consider the Landeau-Zener theory of level crossings \cite{QA}. During
an avoided crossing, the time needed in order to actually reach the
ground state is bounded by the inverse of the energy gap $\Delta$
between these states. If the total annealing is longer $\tau \gg
\Delta^{-2}$, then the adiabtic theorem imply that at each time step,
the systems remains in the ground state. Otherwise, the system misses
the crossing and is not in the ground state at the end of the
computation. A good estimates of the running time of the algoritmh is
thus obtain by the minimal energy gap $\Delta_{min}$ during the
annealing process \cite{QA}.

We will see that mean-field first-order transitions have generically
an exponentially small gap $\Delta \propto N e^{-\alpha N}$ where $N$
is the system size. This implies $\tau \gg e^N$, that is to say:
quantum annealing is an exponentially slow algorithm for a mean-field
system with a first-order transition\footnote{In finite dimensions one
  expects that nucleation will help. However, optimization problems
  are not finite dimensional generically.}.

The goal of this Letter is to illustrate these features via a complete
analytical and detailed numerical analysis for a family of models. We
consider the ferromagnetic $p$-spin model, which reduces to a
mean-field ferromagnet for the case $p=2$ and to the Grover problem
when $p \rightarrow \infty$. We show how to solve the thermodynamics
of these models using standard tools of statistical physics. We
perform extensive numerical studies of the gap for the case of $p$
finite and odd. By introducing an ansatz for the dominant instantonic
pathways, we show how to compute the coefficient in the exponential
decay of the gap.

\section{The simplest quantum ferromagnet}
\label{sec:simplest-ferro}
We consider a Hamiltonian with $N$ Pauli spins $\s$ of the form $
{\cal H}={\cal H}_{z}+\Gamma V$ where ${\cal H}_{z}$ is a function of
the longitudinal values $\sigma^z$ of the spins. ${\cal H}_{z}$ is
thus diagonal in the $\s^z$ representation. We focus on the
ferromagnetic $p$-spin model:
\bea &{\cal H}&=-\frac{1}{N^{p-1}} \sum_{i_1,\ldots,i_p}
\s^z_{i_1}\ldots \s^z_{i_p} - \Gamma \sum_i \s_i^x 
\label{eq:Hamiltonian-ferro}
\\
&=& \!\!\!\!\!\!\!-\frac{M^p(\vec{\s^z})}{N^{p-1}} - \Gamma
M^T(\vec{\s^x}) = \!\! -N \left[ m^p(\vec{\s^z}) - \Gamma
  m^T(\vec{\s^x}) \right] \nonumber
\eea
where we have defined the longitudinal magnetization
$M(\vec{\s^z})=\sum_i \s_i^z$ and the transverse one
$M^T(\vec{\s^x})=\sum_i \s_i^x$ and their magnetization by site
$m=M/N$ and $m^T=M^T/N$. That sort of models were introduced initially
in a spin-glass context in \cite{PSPIN,MPV}. The ground state of
the classical problem, when $\Gamma=0$,
corresponds to all spins aligned in the same direction. Whereas both
the {\it up} and {\it down} states are valid ground states for even
$p$, the {\it up} state is the unique ground state for odd $p$, and we
will concentrate on this case for simplicity. The case $p=2$ is the
usual Curie-Weiss model, where the transition is continuous
\cite{QCW,QuantumCavity}. For $p>2$ however, both quantum and thermal
transitions are discontinuous.  Of special interest is the limit $p
\to \infty$ where for $p$ odd $m(\{{\vec{S}\}})^p \to \pm 1$ if $m=\pm
1$, and zero otherwise. It leads to:
\be {\cal H}= -N \I\((\sum_i \s^z= N\)) + \Gamma \sum_i \s^x
\label{eq:Hamiltonian-simpleferro}
\ee
where the function $\I(x)$ is $1$ if $x$ is true and zero otherwise.
We now specialize to this $p=\infty$ limit.
\section{The $p=\infty$ limit}
\subsection{The classical case: $\Gamma=0$} The $p=\infty$ model is
trivial in the limit $\Gamma \to 0$ where there are only two levels
with nonzero energies $E=N$ and $E=-N$. The partition sum is thus
$Z=2^{N}-2+2\cosh{\beta N}$ so that
\bea \nonumber f &=&\lim_{N \to \infty} -\frac 1{\beta N} \log{\((
  2\cosh{\beta N} + 2^N -2 \))} \\ \nonumber &\approx& \lim_{N \to
  \infty} - \frac 1{\beta N} \log{\((
  e^{\beta N} \(( 1 + e^{N \((\log{2}-\beta\))}\))\))} \\
&=& \min{(f_P,f_P)} ~\tx{with} ~~~~~f_F=-1 ~\tx{and}~
f_P=-\frac{\log{2}}{\beta} \nonumber \eea
One recognizes a first-order transition at $\beta_c =\log 2$ between
two phases that are always locally stable (no spinodal): a
ferromagnetic phase that consists of the classical configuration where
all spins are up for $\beta>\beta_c$ and a trivial paramagnetic phase
at larger temperature.

\subsection{The extreme quantum case: $\Gamma=\infty$}
When $\Gamma$ is large the classical part of $\cal{H}$ can be
neglected; we then find, in the $\s^x$ basis, $N$ independent
classical spins in a field $\Gamma$:
\be f_{QP}=-T \log{2}-T\log{\((\cosh{\Gamma/T}\))}. \ee
The entropy density is given by the logarithm of a binomial in
$[-\Gamma N,\Gamma N]$: this is a perfect quantum paramagnet.
\subsection{The general case}
For $\Gamma=0$ and inverse temperature $\beta<\log{2}$ we saw that the
classical model is just a model where (almost) all configurations have
zero energy. In this case, we thus can ignore the two nonzero levels
and we expect the quantum paramagnetic free energy $f_{QP}$ to be
valid for all $\Gamma$. A simple perturbation computation -- given in
the next section-- shows that this is true in the low-temperature
phase as well, when $\beta>\log{2}$. The system thus has two distinct
phases, the first a quantum paramagnetic and the second a
ferromagnetic phase. A first-order transition occurs when the free
energies cross so that $f=\min{\((f_{QP},f_{F}\))}$. The phase diagram
of the model is very simple: For low $\Gamma$ and $T$, the free-energy
density is that of the classical model in the ferromagnetic phase,
while for larger $\Gamma$ it jumps to the quantum paramagnetic free
energy; a first-order transition separates the two different behaviors
at the value $\Gamma$ such that $f_{F}=f_{QP}$; this happens on the
line defined by
\be \Gamma= \frac 1{\beta}{\rm arccosh}{\frac {e^{\beta}}2} \ee where
the magnetization jumps from $0$ to $1$ (see Fig.~\ref{fig:phase}).
The zero-temperature behavior can be understood from standard
Rayleigh-Schr\"odinger perturbation theory~\cite{RS}. Consider the set
of eigenvalues $E_k$ and eigenvectors $|k \rangle$ of the unperturbed
model, when $\Gamma=0$.  The series for the lowest perturbed
eigenvalue $E_{min}(\Gamma)$ reads
\be
  E_{min}(\Gamma)=E_{min} + \Gamma V_{ii} + \sum_{k \neq min}
  \frac{\Gamma^2 V_{min~k}V_{k~min} }{E_{min}-E_k} + \ldots \, .
  \label{serie1-f}
\ee
Since $V_{ij} \neq 0$ if and only if the two configuration $i$ and $j$
differ by a single spin flip, odd orders do not contribute in
Eq.~(\ref{serie1-f}).  Noting that $\sum_{k \neq n} |V_{nk}|^2$
reduces to a sum over the $N$ levels connected to $E_i$ by a single
spin flip, and using the fact that all $E_k=0$ (except $E_{min}=-N$),
succesives terms are easyly computed and one finds, to all (finite)
orders (see \cite{US-PRL} for a similar computation):
\be E_{min}(\Gamma)= -N - \Gamma^2 + o\((1\)).\ee

The expansion can also be performed using now $\Gamma V$ as a starting
point and with ${\cal H}_0$ as perturbation. Consider the eigenvalue
$-N\Gamma$. In the base $ \mid N \rangle$ corresponding to the
eigenvalues of $\Gamma V$~\footnote{Note that in the $\s^z$ basis the
  ground-state vector $ \mid N \rangle$ has elements $\pm 2^{N/2}$.},
we obtain
\be E(\Gamma) = -N\Gamma + \langle N \mid {\cal{H}}_0 \mid N \rangle +
\sum_{k \neq n} \frac{|\langle k \mid {\cal{H}}_0\mid N \rangle| ^2
}{-N\Gamma-E_k} + \ldots ~.\ee
Denoting $a(l)$ the elements of the vector $ \mid N \rangle$ in the
$z$ basis, the first-order term in this expansion reads $-N
a^2(1)$. Since the $a(l)$ are of order $2^{-N/2}$ the first-order
shift is tiny.  The next term involves a sum over the $2^N-1$ levels
\be \sum_{k \neq min} \frac{|a(1)k(1)| ^2}{-\Gamma N-E_k} =
\sum_{k \neq min} \frac{2^{-N} |k(1)| ^2}{-\Gamma N-E_k}\, . \ee
The last sum is entropically dominated by the states with $E_k=0$ and
therefore gives a negligible contribution (as one can check term by
term). Subsequent terms are treated similarly. This yields the
ground-state energy:
\bea E_{GS}&=&-N-\Gamma^2 +o(1)~\tx{for}~\Gamma<\Gamma_c \\
E_{GS}&=&-\Gamma N +o(1)~\tx{for}~\Gamma>\Gamma_c  \\
&\tx{with}& \Gamma_c=1 +O(1/N).
\eea
\subsection{Exponential closure of the gap}
Near the transition the treatment must be refined: There is an
(avoided) level crossing at $\Gamma_c=1$ in the large $N$ limit
between the paramagnetic and the ferromagnetic ground state. We now
compute the behavior of the quantum gap around $\Gamma_c=1$. We write
the Hamiltonian in the $\s^x$ basis:
\be {\cal H}_{ij}=\Gamma \eps_i \delta_{i,j} + E_c a_i a_j \ee
where $\vec{a}$ is the state corresponding to all spins aligned in the
$z$ direction expressed in the $x$ basis.  $E_c=-N$ and $\eps_i$s
are the (binomially distributed) energies due to the quantum
interaction. With an appropriate convention for the eigenvectors we can
take for $a$ the vector $2^{-N/2}(1,1,1,....1)$.  In this basis, on
multiplying with an eigenvector $\vec{v}$ of eigenvalue $\lambda$, we
find
\bea 0\!&=&\!\((\Gamma \eps_i-\lambda\)) v_i + E_c a_i (\vec{a}
. \vec{v}) =v_i + E_c \frac{a_i}{\Gamma \eps_i-\lambda} (\vec{a}
. \vec{v}) \nonumber \eea
 Multiplying again by $\vec{a}$, we find \be (\vec{a}
. \vec{v}) + E_c \sum_i \frac {a_i^2 (\vec{a} . \vec{v})}{\Gamma
  \eps_i-\lambda}=0 \ee 
so that \be \frac{N}{2^N} \sum_i \frac
1{\Gamma \eps_i-\lambda}=1 \, .
\label{eq:disper}
\ee

The qualitative behavior of the eigenvalues can now be understood
graphically: Between each pole in the denominator of
Eq.~(\ref{eq:disper}) the function passes from $-\infty$ to $+\infty$
passing through unity. All interior roots to the function are thus
bracketed by a comb of poles separated by $2\Gamma$. In the small
$\Gamma$ phase this rigorously brackets almost all the eigenvalues
near $\lambda=0$.  The exception is the lowest eigenvalue which can split off
from the comb, a sign of the phase transition in the large $N$ limit.

In the paramagnetic phase, the lowest eigenvalue is very close to
$\lambda=-\Gamma N$. In this case $-\Gamma N-\lambda$ is very small so
that we can write $\lambda=-\Gamma N +\eta$. In addition the
overwelming majority of eigenvalues $\epsilon_i$ are close to zero~\footnote{Systematic corrections to this approximation do not change
  the result.}; Eq.~(\ref{eq:disper}) then implies, at the transition
when $\Gamma=1$
\bea \nonumber 1&=&\frac{N}{2^N} \left[ \frac 1{-N - \lambda} +
  \frac{2^N-1}{-\lambda}\right] =\frac{N}{2^N} \left[ - \frac 1{\eta}
  - \frac{2^N-1}{\eta-N}\right], \eea
so that finally $\eta^2= N^2/ 2^N$ at the critical point and
\be \Delta_{min} =  2 N  2^{-N/2} \, .
\label{gap-pinfty}
\ee
The gap closes exponentially fast at the transition.  We have an
extremely simple model with a first-order mean-field transition and
most of the physics discussed in this Letter is already present in
this model: difficult problems, such as this one where only one in
$2^N$ configurations has a low energy, manifest themselves by a
first-order transition in the quantum annealing path, and consequently
by an exponentially small gap.

The reader could at this point argue that we have not shown that {\it
  all choices} of the quantum interaction lead to this result; perhaps
a more intelligent choice would turn the transition to second order, and
make the gap polynomial? We know that for this precise model, this is
just impossible. In fact, this model is nothing else than the Grover
problem \cite{Grover}, that is: searching for a minimum value in an
unsorted database. The best algorithm is known, and it is an
exponential one \cite{Grover}. It is obtained by adjusting the
evolution rate of the Hamiltonian in the quantum annealing process so
as to keep the evolution adiabatic on each infinitesimal time
interval.  In doing so, the total running time can be
$\tau\propto\Delta^{-1} $ \cite{Roland}, which is still exponential. There
is thus no way to avoid the exponential gap in this situation.

\section{Behavior for general $p$}
\label{sec:pspin-ferro}
We now consider finite value of $p$ and begin by calculating the phase
diagram in the static approximation.  We then consider closure of the
gap using numerical diagonalization and an instantonic calculation
which we then compare.

\subsection{Phase diagram}
\begin{figure}[t]
  \includegraphics[width=0.48\textwidth]{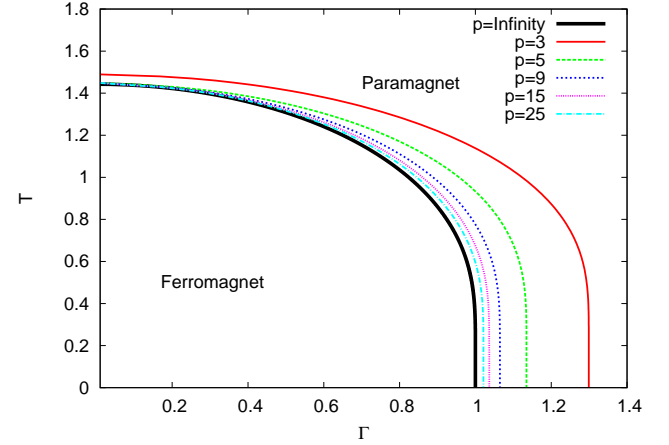}
  \caption{Phase diagram of the ferromagnetic $p$-spin ferromagnet for
    different values of $p$. A first-order transition separates the
    ferromagnetic and quantum paramagnetic phases.}
  \label{fig:phase}
\end{figure}

We shall first use the Suzuki-Trotter formula in order to map onto a
classical model with an additional ``time'' dimension:
\bea Z\!\!\!\!\!&=&\!\!\!\!\!\!\! \sum_{\{{\vec{\s}\}}} \!\!\! \((e^{-\beta
  {\cal H}_{z} + \beta \Gamma \sum_i \s_i^x} \))\!\!\!\!\
=\!\!\!\!\! \lim_{N_s \to \infty}\!\!\!\! \Tr_{\{{\vec{\s}\}}} \!\!\! \left[e^{-\frac{\beta}{N_s} {\cal H}_{z}} e^{\frac{\beta}{N_s} \Gamma \sum_i \s_i^x} \right]^{N_s} \nonumber \\\nonumber
&=& \!\!\!\!\!\!\!\lim_{N_s \to \infty} \sum_{\{{\vec{\s}\}}} \langle
\vec{\s} | e^{-\sum_{\alpha=1}^{N_s}\frac{\beta}{N_s} {\cal
    H}_{z}(\alpha)} e^{\sum_{\alpha=1}^{N_s}\frac{\beta}{N_s} \Gamma
  \sum_i \s_i^x(\alpha)} |\vec{\s} \rangle \, . \eea
We then introduce $N$ closure relations $\I=\sum_{\{{\vec{\s}\}}}
|\vec{\s} \rangle \langle \vec{\s} | $:
\bea \nonumber Z&=& \!\!\!\!\!\!\! \sum_{\{{\vec{\s(\alpha)}\}}}
\prod_{\alpha=1}^{N_s} \langle \vec{\s(\alpha)} |
e^{-\frac{\beta}{N_s} {\cal H}_{z}(\alpha)}
e^{\frac{\beta}{N_s} \Gamma \sum_i \s_i^x(\alpha)} |\vec{\s(\alpha+1)} \rangle \\
\nonumber &=& \!\!\!\!\!\!\!\!\! \sum_{\{{\vec{\s(\alpha)}\}}} \!
\prod_{\alpha=1}^{N_s} e^{-\frac{\beta}{N_s} {\cal H}_{z}(\alpha)} \!
\prod_{\alpha=1}^{N_s} \langle \vec{\s(\alpha)} |e^{\frac{\beta}{N_s}
  \Gamma \sum_i \s_i^x(\alpha)} |\vec{\s(\alpha\!+\!1)}\rangle \eea
with the convention that $\vec{\s(N_s+1)}=\vec{\s(1)}$. Applying $N_s$
times the integral representation of the delta function $ \int dm
\delta(Nm- M(\{{\vec{S}\}})) f(Nm) = f( M(\{{\vec{S}\}}))$, one finds:
\bea \nonumber \!\!\!\!\!\!\!\! &Z&=\!\!\! \int
\prod_{\alpha=1}^{N_s}dm(\alpha) \prod_{\alpha=1}^{N_s}
d\lambda(\alpha) \exp{\(( \frac{\beta N}{N_s} \sum_{\alpha=1}^{N_s}
  m(\alpha)^p \))} \times \\ \nonumber \!\!\!\!\!\!\!\!
&\exp& \!\!\!\!\! {\left[ \!\! \frac{-N}{N_s} \sum_{\alpha=1}^{N_s}
\!\!  \lambda(\alpha) m(\alpha) \!+\! N \log{\Tr\! \prod_{\alpha=1}^{N_s}
    e^{\left[ \frac{\beta}{N_s} \Gamma \s^x(\alpha) +
        \frac{\lambda(\alpha)}{N_s} \s^z(\alpha) \right]}} \!\! \right]}\!. \eea
The saddle point condition imposes that $\lambda(\alpha)=\beta
pm^{p-1}(\alpha)$. Writing $t=\beta \alpha/N_s$ and performing the
limit $N_s \to \infty$ we obtain:
\be Z\!\! =\!\!\! \int \!\!{\cal D}m(t) e^{N \!\int_0^{\beta} dt (1-p) m^p(t) +
  N\! \log{\!\Tr~\!e^{\int_0^{\beta} dt \Gamma \s^x(t) +
        pm^{p-1}\!(t)\! \s^z(t)}}}\!\!. 
\label{voila}
\ee
We now use the ``static'' approximation, which we also check
numerically
\cite{QCW,QuantumCavity}, and remove all ``time'' indices for $m$ to
finally obtain:
\bea \!\!&Z& = \int dm e^{ -\beta N f(\beta,\Gamma,m)} \\ \nonumber
\!\!&f(\beta,\Gamma,m)&\!\!\!\!=\!(p-1) m^p \!-\! \frac 1{\beta}
\log{2\cosh{\((\beta \sqrt{\Gamma^2 + p^2m ^{2p-2}} \))}}.
\label{freeNRG}
\eea
All thermodynamic quantities can now be computed. For instance, the
self-consistent equation for the magnetization $m$ reads (for $p>2$)
\be
m=\((\frac{\tanh{\((\beta\sqrt{\Gamma^2+p^2m^{2p-2}}\))}}{\sqrt{\Gamma^2+p^2m^{2p-2}}}\))
p m^{p-1}.
\label{self:m}
\ee
It is easy to check that the former expression leads to first-order
(quantum and classical) transitions when its minima cross. In
particular, the free energy for $p\! \to\! \infty$ is simply $f\!=\!-1$ for
$m\!=\!1$ and $f=\! - \frac 1{\beta} \log{2\cosh{\((\beta \Gamma \))}}$
otherwise, as we obtained in the first section. The phase diagram of
the model is plotted in Fig.~\ref{fig:phase}. 

The energy is given by $e=\frac {\partial}{\partial \beta} \beta f$,
and thus at low $T$:
\bea \!\!\!
&\e(\beta)& \approx 
 e_{GS} + 2 \sqrt{\Gamma^2 + p^2m
  ^{2p-2}} e^{-2\beta \sqrt{\Gamma^2 + p^2m ^{2p-2}}} \nonumber \\ \nonumber
&\tx{with}& e_{GS}(\Gamma,m)=(p-1) m^p - \sqrt{\Gamma^2 + p^2m ^{2p-2}} .
\eea
In the low-temperature $T$, the energy of a system with $N$ excited
states with an energy gap $\Delta E$ is $E=E_{GS}+N\Delta E e^{-\beta
  \Delta E}$, and this computation thus shows that there are $N$
levels with an energy gap $\Delta E=2\sqrt{\Gamma^2 + p^2m ^{2p-2}}$
where $\Delta E$ is discontinuous at the transition.

This is, however, only a crude description of the phenomenology of the
low-lying states. If indeed only {\it one} level is closing at the
transition, then we expect the energy to behave as $E=E_{GS}+ \Delta E
e^{-\beta \Delta E}$, and therefore one needs to compute the $O(1)$
correction to the energy in order to take this into account. The
former computation thus misses this behavior and indeed, numerical
results show that the first excited state is unique. Worse, we expect
the energy gap between the ground state and the excited state to close
exponentially fast at the transition, and therefore, in order to be
able to investigate this behavior, we should be looking for an
exponentially small gap: in that case we thus need to look for
exponentially small correction to the free energy! Fortunatly, there
is a way to deal with this problem: we now turn to a numerical study
of the gap and to the instantonic approach.

\section{Closure of the gap}

\subsection{Numerical methods}
We use two complementary methods to study the spectrum of the $p$-spin
model for $3\leq p \leq31$.  The full matrix representation of the Hamiltonian
is a sparse operator of dimension $2^N$. For such sparse matrices
Laczos methods are particularly useful and can be used to extract
extremal eigenvalues from the spectrum for $N \le 21$.  We note in
particular that that for $N\le 21$ the transition occurs between two
states with the maximum possible angular momentum $l=N/2$.

\begin{figure}[ht]
  \includegraphics[width=0.48\textwidth]{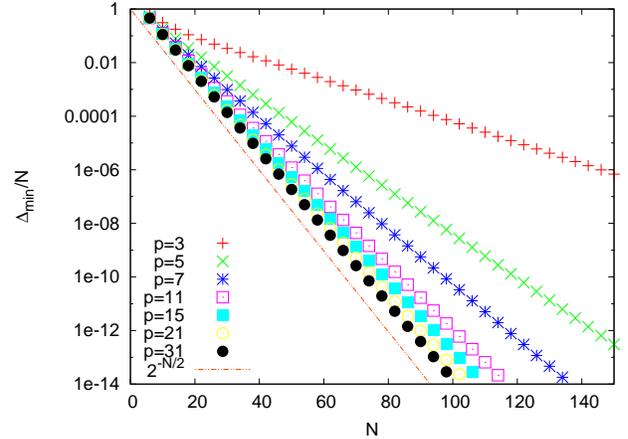}
  \caption{Minimum Gap versus N from exact diagonalization of the
    ferromagnetic p-spin model for some values of $p$ on a linear-log
    scale. One clearly sees that for each $p$ the gap closes
    exponentially with $N$, so that $\Delta_{min} \propto N 2^{-N
      \alpha}$. }
  \label{fig:diag_min}
\end{figure}

Considerable improvements in efficiency are obtained by realazing that
the total angular momentum $L^2$ commutes with ${\cal H}$.  Thus
the transition occurs in a subspace of dimension $2l+1=N+1$. In this
subspace the Hamiltonian has diagonal elements corresponding to
different values of $L^z$. Standard methods from the theory of angular
momentum show that the off-diagonal elements of the matrix in this
subspace are only those labeled by $(m_z,m_z\pm 1)$. The matrix is
symmetric with off-diagonal elements
\begin{equation}
  H_{m_z,m_z+1} = {\Gamma} \sqrt{l(l+1)-m_z(m_z+1)}
\end{equation}
The resulting tri-diagonal matrix an be treated with very high
efficiency allowing one to study systems of $N\sim 100$ in just a few
seconds. The limiting factor in the study of even larger systems is
the reduction of the gap to double precision machine accuracy so that
floating point round-off errors dominate the results.
Fig.~\ref{fig:diag_min} shows the dependence of the minimum gap for
some values of $p$. We see that for all $p\ge 3$ the gap closes exponentially in $N$.

\begin{figure}
  \includegraphics[width=0.48\textwidth]{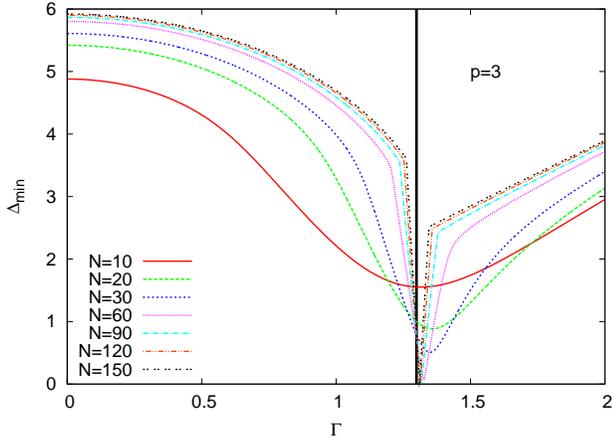}
  \caption{Numerical computation of the gap versus $\Gamma$ for $p=3$
    computed using the method described in the text.  Very 
    close to the transition at $\Gamma_c$ (the black vertical line), in a region that shrinks as $N$ increases,  the gap is closing exponentially fast.
    \label{fig:gaps}}
\end{figure}

Fig.~\ref{fig:gaps} shows the dependence of the gap $\Delta$ as a
function function of $\Gamma$ for $p=3$ and different $N$. $\Delta$
indeed closes fast at the transition that arises exactly at the
critical value predicted analytically. The region where the gap closes
is getting narrow as $N$ increases, and one has to be very careful in
scanning $\Gamma$ in order not to miss it: this is an important
message for future numerical simulations.
Fig.~\ref{fig:diag_min} shows the dependence of the minimum value of
the gap $\Delta_{min}$ as a function of $N$ for some values of
$p$. For all $p\ge 3$, the gap decays exponentially as $\Delta_{min}
\propto N 2^{-N \alpha}$. The different values of $\alpha$ are given
in Table~\ref{tab:ferro_data}. As we expected, the gap closes
exponentially fast at the first-order transition point.  We want now
to show how the coefficient in the exponent can be computed
analytically.



\subsection{The Instantonic approach}
It is well known that the tunneling between quantum states can be
computed using an instantonic approach \cite{Zinn}. Let us briefly
explain how this can be understood via corrections to the saddle-point
computation. At the transition, two solutions (the ferromagnetic one
$m=m_{eq}$ and the paramagnetic one $m=0$) have the same free energies
$f_m=f_{QP}$. Let us assume now that we are able to find another
time-dependent path $m(t)$ ---which we shall call instantonic--- that
spends some time $\tau_1$ in the ferromagnetic state and then jumps to
the paramagnetic state where it spends a time $\tau_2$, and that
exactly at the transition, one has $\epsilon=e^{-N \beta
  \((f_{inst}-f_{ferro}\))}=e^{-N G}$ with $G=O(1)$ in the zero-temperature 
limit. Since we are summing over all periodic paths, one
should now take into account all such instantonic paths that jump an
even number of times to compute the correction to
Eq.~(\ref{freeNRG}). Each of these jumps can occur at any time $t \in
[0,\beta]$ and the saddle-point computation thus reads, at the
transition:
\bea Z &=& 2 e^{-\beta F} + 2 e^{-\beta F} \(( \frac {\beta^2}2
\epsilon^2 + \frac
{\beta^4}{4!} \epsilon^4 + \frac{\beta^6}{6!} + \ldots \)) \nonumber \\
&=& 2 \sum_{k~{\rm even}} \frac{\beta^k}{k!} e^{-\beta F} {\eps}^{k},
\eea
where the factor $\beta^k/k!$ comes from the counting of all possible
paths with $k$ jumps. One then recognizes the series expansion of an
effective two level system:
\be Z=\Tr~ e^{-\beta {\cal H}_{eff}},~{\rm with} \quad {\cal H}_{eff} =
\left( \begin{array}{cc}
    F & \epsilon  \\
    \epsilon & F \end{array} \right) . \ee
Diagonalizing the effective Hamiltonian at $T=0$ one sees immediately
that the gap goes as $\Delta \propto \epsilon=e^{-N G}$: the energy
cost of the instanton thus provides the exponent of the gap at the
transition.

\subsection{Computing the Instanton}
\begin{table}
  \begin{center}
    \begin{tabular}{lcc|c|c|c}
      \hline\noalign{\smallskip}
      $p$ & $\Gamma_c$ & $m_c$ & $\alpha^{Gap}_{sharp}$ & $\alpha^{Gap}_{tanh}$ & $\alpha^{Gap}_{simu}$ \\
      \noalign{\smallskip}\hline\noalign{\smallskip}
      3  & 1.2991   &    0.8660 &     0.2075     &   0.1251    &  0.126(3)      \\
      5  & 1.1347   &    0.9682  &    0.3390    &   0.2686   &  0.270(3)   \\
      7  & 1.0874   &    0.9860  &    0.3888     &   0.3335   &  0.335(3)     \\
      9  & 1.0647   &    0.9921  &    0.4150     &   0.3699   &  0.370(3)    \\
      11 & 1.0514   &   0.9959   & 0.4318      &   0.3929   &  0.395(3)     \\
      13 & 1.0426   &   0.9965   & 0.4422     &   0.4105   &  0.410(3)        \\
      15 & 1.0364 & 0.9974   & 0.4502      &   0.4224 &   0.421(3)      \\
      17 & 1.0318 & 0.9980   & 0.4564        &   0.4315 &   0.431(3)     \\
      19 & 1.0282 & 0.9985   & 0.4620        &   0.4387 &     0.439(3)     \\
      21 & 1.0253 & 0.9987   & 0.4648        &   0.4445         &     0.445(3)    \\
      23 & 1.0230 & 0.9990   & 0.4679        &   0.4493 &     0.450(3)     \\
      25 & 1.0211 & 0.9991   & 0.4705        &   0.4534   &     0.454(3)    \\
      31 & 1.0168 &   0.9994  &   0.4763     &   0.4623   &   0.462(3)   \\
      $\ldots$ &  $1+\frac 1{2p}$ & $1-\frac 1{2p^2}$ & $\frac 12 - \frac {\log{2}}{p}$ &  \ldots & $\frac 12 -\frac {1.15...}p$ \\
      $\infty$ & 1  & 1  & $\frac 12$  & $\frac 12$ & $\frac 12$ \\
      \noalign{\smallskip}\hline
    \end{tabular}
\caption{First-order transition in the p-spin ferromagnet at zero
  temperature: The critical values for the field $\Gamma_c$ and
  magnetization $m_c$ are given. The gap at the transition decays
  exponentially fast as $\Delta \propto N 2^{-N\alpha^{Gap}}$ and we
  give the numerical results from exact diagonalization
  $\alpha^{Gap}_{simu}$, the estimates with the sharp instanton
  $\alpha^{Gap}_{sharp}$ (an upper bound on the true value) and the
  soft instanton $\alpha^{Gap}_{tanh}$: these values are indistinguishable 
  from the numerical ones.
\label{tab:ferro_data}  
}
\end{center}
\end{table}
We can consider various ans\"atze to compute the optimal instanton,
all of them giving lower bounds on the coefficient. The simplest one
is just a sharp wall when $m(t)$ jumps abruptly from the value $m_Q$
to $m_F$.  The gap thus reads in this approximation:\footnote{This can
  be seen in the discrete Suzuki-Trotter formalism where
  \be Z = \sum_{\{{\vec{\s(\alpha)}\}}} \prod_{\alpha=1}^{N_s} \langle
  \vec{\s(\alpha)} |e^{\frac{\beta}{N_s} \(( \Gamma \s^x(\alpha) + h
    \s^z \)) } |\vec{\s(\alpha+1)}\rangle \, . \ee
  Each term but one can be written in its respective diagonal base
  $(1)$ or $(2)$ and be computed with the static approach. However,
  there is a remaining term of the form
  \be \langle \vec{\s_1} |e^{\frac{\beta}{N_s} \(( \Gamma \s^x(\alpha)
    + h \s^z \)) } |\vec{\s_2}\rangle = \langle \vec{\s_1}
  |\vec{\s_2}\rangle \langle \vec{\s_1} |e^{\frac{\beta}{N_s} \((
    \Gamma \s^x(\alpha) + h \s^z \))} |\vec{\s_1}\rangle \ee
}
\be \Delta=\langle F | Q \rangle ^N = e^{N\log{\langle F | Q \rangle}}
\ee
where $\langle F |$ and $\langle Q |$ are the eigenvectors of the matrix
\be \left( \begin{array}{cc}
    p m^{p-1} & \Gamma  \\
    \Gamma & -p m^{p-1}  \\
  \end{array} \right)
\ee
%
%
%
Exactly at the transition, this gives an estimates on the gap
$\Delta$. In particular for $p \to \infty$, we find that $\Delta
\approx N 2^{-N/2}$, as was previously found in the first section. For
finite $p$, however this yields only a crude lower bound on the value
of the exponent (see Table~\ref{tab:ferro_data}).

We thus use a $tanh$ shape for $m(t)$ and compute numerically the
cost, by integrating Eq.(\ref{voila}). We use the width of the $tanh$
function as a variational parameter which we vary in order to minimize
the estimate of the instanton free energy from which we deduce the
gap. The results of this procedure are given in
Table~\ref{tab:ferro_data}. When we now compare the numerical data
from exact diagonalization with the prediction from the instantonic
computation, we observe that there is {\it no detectable difference
  within our numerical precision} between the instantonic prediction
from the $tanh$ shape and the numerical estimation of the
coefficient. We have thus obtained $\Delta$ from first-principle
computations.

\section{Conclusions}
Quantum annealing has been presented as a new way of solving hard
optimization problems with complicated and rough configuration
spaces. In this paper we have shown that even in systems with trivial
energy landscapes quantum annealing can fail (and there is thus no
need for more complex phenomena to explain this failure, as for
instance in \cite{BOF}). Already the $p=3$ ferromagnet exhibits a
first-order phase transition with an exponentially closing gap: A
scenario which is very pessimistic for the success of the quantum
annealing algorithm.  We have also shown that the $p=\infty$ limit of
the ferromagnetic model is related to the Grover problem. This is a
clear indication that these first-order transition carry the signature
of the most difficult problems.

Models presented in this Letter allow a complete analytical and
numerical treatment.  Their disordered counterpart can be studied
using the generalized instanton introduced in
\cite{US-PRL,US-Long}. It would be interesting to extend this approach
to dilute mean-field system and random optimization problems, using
the quantum cavity of \cite{QuantumCavity,XORSAT}.

\begin{acknowledgements}
  We thank P.~Boniface, S.~Franz, A.~Rosso, G.~Semerjian,
  L.~Zdeborov\'a and F.~Zamponi for discussions.
\end{acknowledgements}

\bibstyle{eplbib}

\end{document}